\documentclass[twocolumn,twocolappendix]{aastex63}
\usepackage{fp}
\usepackage{bm}
\usepackage{xcolor}
\DeclareSymbolFont{matha}{OML}{txmi}{m}{it}
\DeclareMathSymbol{\varv}{\mathord}{matha}{118}

\usepackage{multirow}
\usepackage{bm}
\usepackage{amsmath}
\usepackage{CJKutf8}
\usepackage{soul}

\newcommand{\N}[1]{\textcolor{blue}{ #1}} 
\definecolor{jaredcolor}{HTML}{5D3FD3}

\shorttitle{Degeneracies in SN~2023ixf's Light-Curve Models}
\shortauthors{Hsu et al.}
\newcommand{\LCO}{\affiliation{Las Cumbres Observatory, 6740 Cortona Drive, Suite 102, Goleta, CA 93117-5575, USA}}
\newcommand{\UCSB}{\affiliation{Department of Physics, University of California, Santa Barbara, CA 93106-9530, USA}}

\newcommand{\Tsinghua}{\affiliation{Physics Department and Tsinghua Center for Astrophysics, Tsinghua University, Beijing, 100084, People's Republic of China}}

\newcommand{\CfA}{\affiliation{Center for Astrophysics \textbar{} Harvard \& Smithsonian, 60 Garden Street, Cambridge, MA 02138-1516, USA}}
\newcommand{\UA}{\affiliation{Steward Observatory, University of Arizona, 933 North Cherry Avenue, Tucson, AZ 85721-0065, USA}}

\newcommand{\IAIFI}{\affiliation{The NSF AI Institute for Artificial Intelligence and Fundamental Interactions}}

\newcommand{\Flatiron}{\affiliation{Center for Computational Astrophysics, Flatiron Institute, 162 5th Avenue, New York, NY 10010-5902, USA}}
\newcommand{\GeminiNorth}{\affiliation{Gemini Observatory, 670 North A`ohoku Place, Hilo, HI 96720-2700, USA}}

\newcommand{\ICE}{\affiliation{Institute of Space Sciences (ICE, CSIC), Campus UAB, Carrer
de Can Magrans, s/n, E-08193 Barcelona, Spain}}
\newcommand{\IEEC}{\affiliation{Institut d'Estudis Espacials de Catalunya (IEEC), Edifici RDIT, Campus UPC, 08860 Castelldefels (Barcelona), Spain}}
\newcommand{\konkoly}{\affiliation{Konkoly Observatory, HUN-REN Research Center for Astronomy and Earth Sciences, Konkoly Th. M. út 15-17., Budapest, 1121 Hungary; MTA Centre of Excellence}}
\newcommand{\szeged}{\affiliation{Department of Experimental Physics, Institute of Physics, University of Szeged, D\'om t\'er 9, Szeged, 6720 Hungary}}
\newcommand{\UCSD}{\affiliation{Department of Astronomy \& Astrophysics, University of California, San Diego, 9500 Gilman Drive, MC 0424, La Jolla, CA 92093-0424, USA}}
\newcommand{\Liverpool}{\affiliation{Astrophysics Research Institute, Liverpool John Moores University, 146 Brownlow Hill, Liverpool, L3 5RF, UK}}

\defcitealias{sdss_2017}{SDSS Collaboration 2017}

\begin{document}

\title{One Year of SN~2023ixf: Breaking Through the Degenerate Parameter Space in Light-Curve Models with Pulsating Progenitors}

\correspondingauthor{Brian Hsu}
\email{bhsu@arizona.edu}

\author[0000-0002-9454-1742]{Brian~Hsu}
\UA

\author[0000-0001-5510-2424]{Nathan~Smith}
\UA

\author[0000-0003-1012-3031]{Jared~A.~Goldberg}
\Flatiron

\author[0000-0002-4924-444X]{K.~Azalee~Bostroem}
\altaffiliation{LSST-DA Catalyst Fellow.}
\UA

\author[0000-0002-0832-2974]{Griffin~Hosseinzadeh}
\UCSD

\author[0000-0003-4102-380X]{David~J.~Sand}
\UA

\author[0000-0002-0744-0047]{Jeniveve~Pearson}
\UA

\author[0000-0002-1125-9187]{Daichi~Hiramatsu}
\CfA\IAIFI

\author[0000-0003-0123-0062]{Jennifer~E.~Andrews}
\GeminiNorth

\author[0000-0003-4666-4606]{Emma~R.~Beasor}
\Liverpool

\author[0000-0002-7937-6371]{Yize Dong \begin{CJK*}{UTF8}{gbsn}(董一泽)\end{CJK*}}
\CfA\IAIFI

\author[0000-0003-4914-5625]{Joseph~Farah}
\LCO\UCSB

\author[0000-0002-1296-6887]{Llu\'{i}s~Galbany}
\ICE\IEEC

\author[0000-0001-6395-6702]{Sebastian~Gomez}
\CfA

\author[0000-0003-0209-9246]{Estefania~Padilla~Gonzalez}
\LCO\UCSB

\author[0000-0003-2375-2064]{Claudia~P.~Guti\'{e}rrez}
\IEEC\ICE

\author[0000-0003-4253-656X]{D.~Andrew~Howell}
\LCO\UCSB

\author[0000-0002-8770-6764]{R\'{e}ka~K\"{o}nyves-T\'{o}th}
\konkoly\szeged

\author[0000-0001-5807-7893]{Curtis~McCully}
\LCO

\author[0000-0001-9570-0584]{Megan~Newsome}
\LCO\UCSB

\author[0000-0002-4022-1874]{Manisha~Shrestha}
\UA

\author[0000-0003-0794-5982]{Giacomo~Terreran}
\LCO

\author[0000-0002-5814-4061]{V.~Ashley~Villar}
\CfA\IAIFI

\author[0000-0002-7334-2357]{Xiaofeng Wang}
\Tsinghua

\begin{abstract}
We present and analyze the extensive optical broadband photometry of the Type II SN~2023ixf up to one year after explosion. We find that, when compared to two pre-existing model grids, the bolometric light curve is consistent with drastically different combinations of progenitor and explosion properties. This may be an effect of known degeneracies in Type IIP light-curve models. We independently compute a large grid of {\tt MESA+STELLA} single-star progenitor and light-curve models with various zero-age main-sequence masses, mass-loss efficiencies, and convective efficiencies. Using the observed progenitor variability as an additional constraint, we select stellar models consistent with the pulsation period and explode them according to previously established scaling laws to match plateau properties. Our hydrodynamic modeling indicates that SN~2023ixf is most consistent with a moderate-energy ($E_{\rm exp}\approx7\times10^{50}$ erg) explosion of an initially high-mass red supergiant progenitor ($\gtrsim 16.5\ M_{\odot}$) that lost a significant amount of mass in its prior evolution, leaving a low-mass hydrogen envelope ($\lesssim 3\ M_{\odot}$) at the time of explosion, with a radius $\gtrsim 950\ R_{\odot}$ and a synthesized $^{56}$Ni mass of $\approx0.068\ M_{\odot}$. We posit that previous mass transfer in a binary system may have stripped the envelope of SN~2023ixf's progenitor. The analysis method with pulsation period presented in this work offers a way to break degeneracies in light-curve modeling in the future, particularly with the upcoming Vera C.~Rubin Observatory Legacy Survey of Space and Time, when a record of progenitor variability will be more common. 
\end{abstract}

\section{Introduction}
\label{sec:intro}

Type II supernovae (SNe II) are hydrogen-rich core-collapse supernovae (CCSNe) that are thought to mark the violent deaths of massive stars ($\gtrsim 8\ M_{\odot}$; \citealt{Woosley_Weaver_1986}). Those with a plateau shape in the light curves (SNe IIP) are the most common variety of massive star explosion, representing about half of all CCSNe (\citealt{Smith_2011}). Their progenitors have been confirmed to be red supergiants (RSGs) by pre-explosion detections, with zero-age main-sequence (ZAMS) masses in the range of $M_{\rm ZAMS}\approx8-20\ M_{\odot}$ (e.g., \citealt{Van_Dyk_2003,Van_Dyk_2012,Van_Dyk_2023,Li_2006,Smartt_2009,Smartt_2015,Davies_Beasor_2020}). Even with direct detections of the progenitor star in pre-explosion imaging, inferring the initial mass of the exploding star is subject to significant uncertainty, leading to much debate about the initial masses of SNe IIP (\citealt{Smartt_2015,Davies_Beasor_2018,Davies_Beasor_2020}). Clues to the star's mass may also be inferred from the shape of the plateau light curve, since this shape is mediated by the recombination of the hydrogen envelope (\citealt{Kirshner_Kwan_1975,Shigeyama_Nomoto_1990,Chugai_1991,Popov_1993,Utrobin_Chugai_2005}). This is complicated, however, by the fact that the light-curve shape depends on other factors like the explosion energy and progenitor radius (\citealt{Litvinova_Nadezhin_1983,Chugai_1991,Kasen_Woosley_2009,Dessart_2013,Pejcha_2015,Sukhbold_2016,Goldberg_2019,Martinez_2022b}).

Extensive photometric and spectroscopic follow-up at multiple wavelengths over the duration of SN II light curves can help probe the underlying physics during different phases of their evolution (e.g., \citealt{Fransson_1984,Dessart_Hillier_2005,Dessart_Hillier_2006,Dessart_Hillier_2008,Dessart_2013,Dessart_2024b,Hillier_2019,Jencson_2019,Kozyreva_2019,Kozyreva_2020,Dessart_Jacobson-Galan_2023}). In particular, when SNe II are discovered very soon after explosion, early observations offer unique insight into explosion geometry, surrounding environments, and mass loss during the final years of a massive star's life (e.g., \citealt{Baron_2000,Brown_2007,Smith_2014,Dessart_Hillier_2022,Jacobson-Galan_2022,Jacobson-Galan_2024,Bostroem_2023b,Bostroem_2023,Li_2024,Shrestha_2024b,Shrestha_2024a}).

Spectroscopic observations of some SNe II within a few days from explosion revealed narrow emission features from slow circumstellar material (CSM) with high-ionization states (e.g., \citealt{niemala85,gh94,quimby07,Gal-Yam_2014,groh14,smith11iqb,shivvers15,Khazov_2016,Yaron_2017,bullivant18,Tartaglia_2021,Bruch_2021,Bruch_2023,Zhang_2023, Andrews_2024,Shrestha_2024b,Shrestha_2024a}). Several SNe II also exhibited light-curve excess above the canonical shock-cooling model (e.g., \citealt{Hosseinzadeh_2018,Andrews_2019,Dong_2021,Tartaglia_2021,Hosseinzadeh_2022,Pearson_2023,Andrews_2024, Shrestha_2024b, Shrestha_2024a}). Combined with recent efforts of detailed hydrodynamic and radiative transfer modeling involving dense CSM (\citealt{Morozova_2017,Morozova_2018,Boian_2019,Dessart_Hillier_2022,Dessart_Jacobson-Galan_2023,Moriya_2023}), observations and theory alike seem to suggest that a significant portion of SN II progenitors were enshrouded in CSM shells or inflated envelopes (e.g., \citealt{Forster_2018,Morozova_2018,Bruch_2023,Jacobson-Galan_2024}). If unbound and sustained, the formation of this dense CSM requires an enhanced mass-loss rate $\sim10^{-4}-10^{-2}\ M_{\odot}\ {\rm yr^{-1}}$ in the months to years prior to the onset of core-collapse, which is much higher than values expected for normal stellar winds of RSGs (e.g., \citealt{Mauron_Josselin_2011,Beasor_Davies_2018,Beasor_2020,Beasor_Smith_2022}). While the exact nature of such an intense mass-loss episode is still unclear, several mechanisms have been proposed, including, but not limited to, late-phase nuclear burning instabilities (\citealt{Arnett_Meakin_2011,Smith_Arnett_2014,Woosley_Heger_2015}), wave-driven mass loss (\citealt{Quataert_Shiode_2012,Shiode_2013,Shiode_Quataert_2013,Fuller_2017,Wu_Fuller_2021,Wu_Fuller_2022}), binary interaction \citep{Smith_Arnett_2014}, atmospheric shocks \citep{Fuller_2024}, and pulsation-driven superwinds (\citealt{Yoon_Cantiello_2010}).

Recently, SN~2023ixf provided another clear case of an SN II progenitor surrounded by dense CSM structures, since it was discovered very early after explosion (\citealt{Hosseinzadeh_2023,Li_2024}) and showed short-lived narrow emission lines in its early spectra (\citealt{Bostroem_2023,Hiramatsu_2023,Jacobson-Galan_2023,Singh_2024,Smith_2023,Zimmerman_2024}). SN~2023ixf is a SN II (\citealt{Perley_2023}) discovered on 2023 May 19 17:27:15.00 UT (\citealt{Itagaki_2023}) in the Pinwheel galaxy. The earliest observation of this SN can be traced back to about 0.9 day before discovery, corresponding to a phase at about 1 hour after explosion (\citealt{Li_2024}). Its proximity to Earth offers an unprecedented opportunity to study the late-stage evolution of RSGs and physics pertaining to SNe II encoded in its extensively sampled light curve across the electromagnetic spectrum. Analyses of both the early light curve and spectral series from radio to $\gamma$-ray wavelengths all indicated the presence of dense CSM surrounding its RSG progenitor (\citealt{Berger_2023,Bostroem_2023b,Grefenstette_2023,Hiramatsu_2023,Jacobson-Galan_2023,Matthews_2023,Panjkov_2023,Smith_2023,Teja_2023,Yamanaka_2023,Zhang_2023,Chandra_2024, Martinez_2024}). This was further supported by optical spectropolarimetry observations (\citealt{Vasylyev_2023, Shrestha_2024c,Singh_2024}) and the evolution of shock breakout emission (\citealt{Li_2024}), suggesting that either the CSM or the explosion was asymmetric. The inferred mass-loss rate from these studies probed a wide range of $10^{-6}-10^{-2}\ M_{\odot}\ {\rm yr}^{-1}$ at various layers of the CSM, indicating a time-variable mass-loss history.

Using pre-explosion images from the {\it Hubble Space Telescope}, {\it Spitzer Space Telescope}, and various ground-based facilities, the properties of the candidate RSG progenitor of SN~2023ixf have also been estimated. In particular, the ZAMS mass estimated independently from spectral energy density (SED) fitting, comparison with single-star evolutionary tracks, environmental study in the vicinity of the SN, and analysis of infrared (IR) variability collectively yield a large range of $M_{\rm ZAMS}\approx 8-20\ M_{\odot}$ (\citealt{Jencson_2023,Niu_2023,Pledger_Shara_2023,Qin_2023,Soraisam_2023,Van_Dyk_2024,Neustadt_2024,Xiang_2024}). Given the uncertainties in both the progenitor's inferred mass-loss rate and initial mass, it is useful to derive the progenitor properties using an alternative method. For example, \cite{Bersten_2024} compared the bolometric light curve and expansion velocity evolution of SN~2023ixf to hydrodynamic models. They found a model with an initial mass of $12\ M_{\odot}$, an explosion energy of $1.2\times10^{51}\ {\rm erg}$, and a synthesized $^{56}$Ni mass of $0.05\ M_{\odot}$ to be compatible with the observed luminosity evolution. Similar low-mass progenitor models were also favored by \cite{Moriya_2024} and \cite{Singh_2024} based on light-curve fitting, and by \cite{Ferrari_2024} based on nebular-phase spectroscopy.

In this paper, we present new observations and an independent analysis of the densely sampled optical light curves of SN~2023ixf. The optical photometry presented in this paper has a $\sim2$ day cadence in 7 bands, extending up to a year after discovery. We give a description of the observations and the data reduction process in Section~\ref{sec:observations}, followed by the methodologies used to calculate the bolometric light curve, plateau properties, and radioactive tail properties of SN~2023ixf. Section~\ref{sec:model_desc} compares our observations to publicly available hydrodynamic model grids and Section~\ref{sec:rsg_mesa_gyre} presents our independent modeling of the bolometric and multi-band light curves. We additionally include the observed RSG variability to aid model selection, and this reveals evidence for a small H-rich envelope mass and high ZAMS mass for SN~2023ixf's progenitor. As such, we discuss the possible formation channel and implications for the progenitor of SN~2023ixf in Section~\ref{sec:discussion}. Finally, we summarize our findings and draw conclusions in Section~\ref{sec:conclusion}.

\section{Observations and Initial Analysis} 
\label{sec:observations}

\subsection{Photometry and Reductions} 
\label{sec:photometry}

\begin{figure*}[t!]
    \centering
    \includegraphics[width=.95\textwidth]{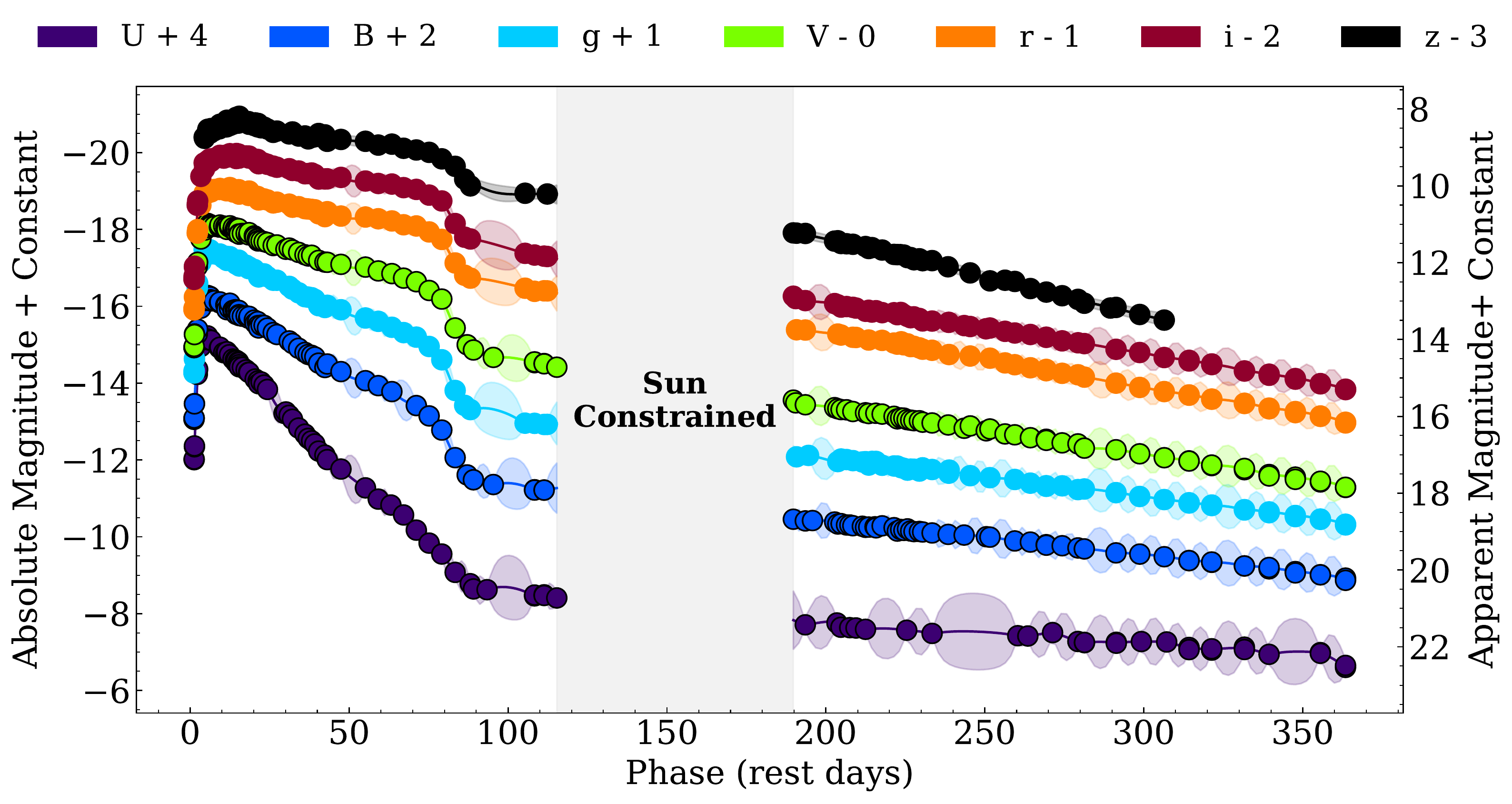}
    \caption{Multi-band light curves of SN~2023ixf from LCO (circles) up to roughly one year after discovery, supplemented with additional UV/optical/NIR photometry from \cite{Singh_2024} (squares) and \cite{Zimmerman_2024} (triangles). All photometry have been corrected for Milky Way and host extinctions. The magnitude uncertainties ($\lesssim 0.1$ mag) are smaller than the marker size.
    \\(The data used to create this figure are available in the published version.)}
    \label{fig:lc}
\end{figure*}

Following the discovery of SN~2023ixf by \cite{Itagaki_2023}, continuous photometric monitoring of the Pinwheel Galaxy was carried out using the Sinistro cameras on Las Cumbres Observatory's robotic 1 m telescopes (\citealt{Brown_2013}) at Teide Observatory (Canary Islands, Spain), McDonald Observatory (Texas, USA), and Haleakalā Observatory (Hawai'i, USA) as part of the Global Supernova Project (GSP) collaboration (\citealt{Howell_2017}). The $UBVgriz_s$ data were reduced using {\tt lcogtsnpipe} (\citealt{Valenti_2016}), a {\tt PyRAF}-based image reduction pipeline that utilizes a standard point-spread function fitting procedure to measure instrumental magnitudes. {\it UBV} magnitudes were calibrated to stars in the L92 standard fields of \cite{Landolt_1983,Landolt_1992} observed on the same night with the same telescopes, {\it gri} magnitudes were calibrated to the AAVSO Photometric All-Sky Survey (\citealt{APASS}) catalog, and {\it z$_{\rm s}$} magnitudes were calibrated to the Sloan Digital Sky Survey (\citetalias{sdss_2017}) catalog. The early photometry up to 2023 June 18 (MJD = 60113; 30 days after discovery) following the same reduction procedure has been presented in \cite{Hosseinzadeh_2023} and \cite{Hiramatsu_2023}\footnote{We independently reduce the early photometry to avoid introducing systematic uncertainties, which may lead to slight differences in the reported magnitudes compared to these previous studies.}. All magnitudes are reported in the AB system and are available as data behind Figure~\ref{fig:lc}.

The Pinwheel Galaxy has a luminosity distance of $d_L=6.71\pm0.14$ Mpc ($\mu=29.135\pm 0.045$ mag; \citealt{Riess_2016}), measured via the \cite{Leavitt_1908} Law of Cepheid variables. The Milky Way reddening in the direction of SN~2023ixf is $E(B-V)_{\rm MW}=0.0077$ mag (\citealt{Schlafly_Finkbeiner_2011}), and the host-galaxy extinction is $E(B-V)_{\rm host}=0.031\pm0.006$ mag (\citealt{Lundquist_2023,Smith_2023}). We correct for both Galactic and host extinction using a \cite{Fitzpatrick_1999} extinction law with $R_V=3.1$ and adopt MJD 60082.788 as the explosion date, following the analysis of \cite{Li_2024}. 

\subsection{Bolometric Light Curve}
\label{sec:pseudo_bol}

Even with the large set of optical photometry covering $3250-9740$ \AA\ presented in this work, \cite{Singh_2024} showed that without ultraviolet (UV) and near-IR (NIR) coverage, the luminosity may be underestimated by a factor of up to $\sim 10$ during the first $\sim 10$ days and a factor of $\sim 2$ during the nebular phase of SN~2023ixf's evolution, respectively. Therefore, to allow for a more robust construction of the bolometric light curve that captures both early- and late-time behaviors, we include NIR and UV photometry from \cite{Singh_2024} and \cite{Zimmerman_2024}. The NIR data contain $J$, $H$, $K_s$ band photometry from various ground-based facilities, while the UV data contain $UVW1$, $UVM2$, $UVW2$ photometry from the Ultraviolet Optical Telescope (UVOT; \citealt{Roming_2005}) onboard the Neil Gehrels Swift Observatory (\citealt{Gehrels_2004}). The inclusion of UV and NIR photometry extends our spectral coverage to $0.16-2.35\ \mu{\rm m}$. To fill in the gap in our LCO light curves from $\sim120-150$ days after explosion, we also supplement our data with optical photometry from \cite{Singh_2024}. All supplemental photometry are shown in Figure~\ref{fig:lc} in different markers.

We first generate a UVOIR pseudo-bolometric light curve of SN~2023ixf up to $\sim 150$ days after explosion, corresponding to the phase where we have complete multi-wavelength coverage from UV to NIR. To obtain observations at all intermediate epochs across all bands, we interpolate the multi-band light curves using a low-order polynomial using {\tt scipy} (\citealt{Virtanen_2020}). Equipped with photometric measurements and interpolations in all available bands at all epochs, we convert the magnitudes to monochromatic fluxes at the mean wavelength of each filter using the transmission functions and magnitude zero-points. The pseudo-bolometric luminosity at each epoch is then calculated via a full integration of the monochromatic fluxes using the trapezoidal rule within the wavelength range. To account for magnitude uncertainties, we perform a Monte Carlo procedure by sampling a Gaussian distribution centered at the magnitude value with the magnitude uncertainty as one standard deviation. This procedure is done 10,000 times for each epoch, and we report the median and standard deviation as the corresponding pseudo-bolometric luminosity and uncertainty.

Next, to account for any potentially missing flux, we perform blackbody (BB) fits to the SEDs of SN~2023ixf to obtain the effective blackbody temperature ($T_{\rm BB}$) and photospheric radius ($R_{\rm BB}$) at each epoch using the {\tt Python}-based MCMC routine {\tt emcee} (\citealt{Foreman-Mackey_2013}), implemented in the Light Curve Fitting package (\citealt{LCFitting}). Here, we follow the same fitting procedure in \cite{Faran_2018} and \cite{Martinez_2022a}, where we remove bands that deviate more than $1\sigma$ from the best-fit BB model, either due to strong line emission (e.g., H$\alpha$ for $r$-band and NIR \ion{Ca}{2} triplet for $z$-band) or line-blanketing effects of iron-group elements. We then calculate additional UV and IR bolometric corrections at each epoch by extrapolating the best-fit BB spectrum from $UW2$-band to $\lambda=0\ {\rm \AA}$ and from $K$-band to infinity, respectively. The full bolometric light curve is then derived by summing the pseudo-bolometric light curve with the bolometric corrections. A similar Monte Carlo procedure is performed by sampling the posterior distributions of the BB fits to obtain uncertainty estimates.

\begin{figure}[t!]
    \centering
    \includegraphics[width=\columnwidth]{BB_params.pdf}
    \caption{Evolution of photospheric temperature and radius of SN~2023ixf estimated from BB fits to the UVOIR multi-band photometry. The region shaded in gray indicates phases where SN~2023ixf's evolution is dominated by energy from CSM interaction (up to $\sim 8$ days after explosion) based on the epoch of disappearance of narrow line features in \cite{Singh_2024}.} 
    \label{fig:BB_params}
\end{figure}

The full evolution of photospheric temperature and radius are shown in Figure~\ref{fig:BB_params}. Similar to previous studies (\citealt{Martinez_2024, Singh_2024,Zimmerman_2024}), we find a steep increase in $T_{\rm BB}$ over the first $\sim 5$ days of SN~2023ixf's evolution. However, the peak temperature of $\approx26,000$ K we find is about $\approx11,000$ K lower. This is due to the fact that the early-time evolution of SN~2023ixf is driven by shock interaction with a dense CSM, which causes its spectrum to depart from a BB. Depending on the combination of bands used (UV+optical vs.~optical+NIR), the inferred temperature can vary up to $\approx 10,000$ K, which translates to a factor of $\sim 2$ in the extrapolated bolometric luminosity. Nevertheless, our temperature evolution is consistent with the spectral BB fits in \cite{Zheng_2025}. We do caution that once SN~2023ixf enters the nebular phase, the ejecta become optically thin and emission dominated, and there may be late-time CSM interaction (\citealt{Bostroem_2024,Kumar_2024,Singh_2024,Folatelli_2025}) in addition to $^{56}$Ni decay. In this case, a BB model cannot accurately capture the continuum flux at longer wavelengths. However, as shown by \cite{Martinez_2022a}, even though BB fits may not be physical during the nebular phase, the bolometric corrections calculated using BB extrapolations are representative of the missing flux at mid-IR wavelengths.

As the goal of this work is to constrain the progenitor and explosion characteristics of SN~2023ixf, which depend primarily on properties during the plateau phase and the radioactive tail, the uncertainty in bolometric luminosity during the CSM-interaction phase should not affect the remainder of our analysis.

\subsection{Measuring Plateau Duration and Nickel Mass}
\label{sec:plateau_fit}

We estimate the plateau duration $t_p$ following \cite{Valenti_2016}\footnote{The second term in the original equation presented in \cite{Valenti_2016} is $P_0\times t$, which is a typo that was confirmed by the corresponding author.}, fitting the functional form $y(t)=V_{\rm mag}$ to the $V$-band light curve and $y(t)=\log_{10}(L_{\rm bol})$ to the bolometric light curve around the fall from the plateau:
\begin{equation}
y(t)=\frac{-A_0}{1+e^{(t-t_p)/W_0}}+P_0\times\left(t- t_p\right)+M_0
\label{eq:y_fit}
\end{equation}

\noindent We fit the light curves starting at day 60 (which corresponds to when the evolution is $\approx$75\% of the way to its steepest descent; \citealt{Goldberg_2019}) to day 150 using the nonlinear least-squares routine {\tt scipy.optimize.curve\_fit}, fixing $P_0$ to be the slope on the $^{56}$Ni tail. The derived median and $1\sigma$ uncertainty for the plateau duration is $t_p=(81.78\pm0.19)$ days for the $V$-band light curve and $t_p=(82.47\pm0.07)$ days for the bolometric light curve, which are consistent with the plateau duration of $t_p=(83.08\pm0.08)$ days in \cite{Bersten_2024}.

Curiously, the derived slope on the radioactive tail for the $V$-band light curve corresponds to a decline rate of $P_0\approx (1.16\pm0.08)$ mag per 100 days\footnote{Note that $P_0$ here, while qualitatively is describing the same portion of the light curve, is not the same as $s_3$ quoted in \cite{Bersten_2024} and \cite{Singh_2024}, which follows the prescription in \cite{Anderson_2014}. If we follow the same procedure, we do recover $s_3=(1.29\pm0.03)$ mag per 100 days using the $V$-band light curve, which is consistent with \cite{Singh_2024}. \cite{Bersten_2024} inferred $s_3=(1.71\pm0.74)$ mag per 100 days by fitting the bolometric magnitudes.}, slightly higher than the value of $\approx 0.98$ mag per 100 days expected for the typical $^{56}$Ni decay. The steepened slope during the nebular phase could be explained by incomplete trapping due to $\gamma$-ray leakage, commonly seen in short-plateau SNe IIP with steeper declines, where the progenitors have partially-stripped, low-density envelopes (e.g., \citealt{Anderson_2014,Morozova_2015,Paxton_2018,Hiramatsu_2021}). In accordance with the steeper decline rate, we fit the bolometric light curve with a modified energy deposition rate from \cite{Wheeler_2015}:
\begin{equation}
L_{\rm tail}=L_{\rm decay}\left[1-e^{-(T_0/t)^2}\right]
\label{eq:Ni_trapping}
\end{equation}

\noindent where 
\begin{equation}
L_{\rm decay}=\frac{M_{\rm Ni}}{M_{\odot}}\left(6.45e^{-t/t_{\rm Ni}} + 1.45e^{-t/t_{\rm Co}}\right)\times10^{43}\ {\rm erg\ s^{-1}}
\label{eq:Ni_lum}
\end{equation}
\noindent is the $^{56}$Ni$\rightarrow$$^{56}$Co$\rightarrow$$^{56}$Fe decay luminosity given by 
\cite{Nadyozhin_1994}, $t_{\rm Ni} = 8.8$ days, $t_{\rm Co} = 111.3$ days, $t$ is the time in days since the explosion, and $T_0$ is the $\gamma$-ray diffusion timescale. Roughly speaking, $T_0$ indicates the strength of $\gamma$-ray trapping within the ejecta. As $T_0\rightarrow\infty$, the instantaneous heating rate given by Equation~\ref{eq:Ni_lum} is recovered and no leakage occurs (i.e., complete trapping). 

Fitting Equation~\ref{eq:Ni_trapping} to our bolometric light curve from day $90-150$ yields a nickel mass of $M_{\rm Ni}=(0.068\pm 0.001)\ M_{\odot}$ and a leakage timescale of $T_0=190\pm3$ days. The $^{56}$Ni mass we derive here is slightly higher than the value of $0.059\ M_{\odot}$ found by \cite{Singh_2024}, marginally consistent with $0.071\ M_{\odot}$ found by \cite{Zimmerman_2024}, and notably higher than $0.04\ M_{\odot}$ found by \cite{Moriya_2024} or $0.05\ M_{\odot}$ found by \cite{Bersten_2024}. The discrepancies in the derived $M_{\rm Ni}$ values may be caused by the timescale used to fit the energy deposition rate (e.g., up to $\sim 110$ days in \citealt{Bersten_2024} and \citealt{Zimmerman_2024}), the spectral range used (e.g., UVOIR in \citealt{Singh_2024} and optical only in \citealt{Moriya_2024}), and procedural difference in calculating bolometric corrections. Other factors such as late-time CSM interaction (\citealt{Singh_2024,Folatelli_2025}), which may contribute up to $5\%$ of the total luminosity (\citealt{Bostroem_2024}), could add to the uncertainty of $M_{\rm Ni}$.  Therefore, the derived value here should be viewed as a rough approximation rather than the true yield of SN~2023ixf.

\section{Comparison to Light-Curve Model Grids}
\label{sec:model_desc}

The main goal of this work is to constrain progenitor properties (e.g., envelope mass and progenitor radius), as well as explosion properties (e.g., explosion energy and nickel mass) for SN~2023ixf. Since these intrinsic SN properties are constrained primarily by quantities on the plateau, we only consider phases when CSM interaction no longer dominates the total luminosity during the plateau evolution, which we take to be $t\geq30$ days based on the break in $R_{\rm BB}$ in Figure~\ref{fig:BB_params} (although there is evidence for low levels of continued interaction---see \citealt{Bostroem_2024} and \citealt{Singh_2024}). In the following section, we compare our bolometric light curve to two sets of model grids (\citealt{Hiramatsu_2021, Moriya_2023}), which vary progenitor, CSM, and explosion properties. For simplicity, we evaluate $\chi^2$ values and infer progenitor and explosion properties based on the $\chi^2$ distribution of each model grid used.

\begin{figure*}[t!]
    \centering
    \includegraphics[width=.95\textwidth]{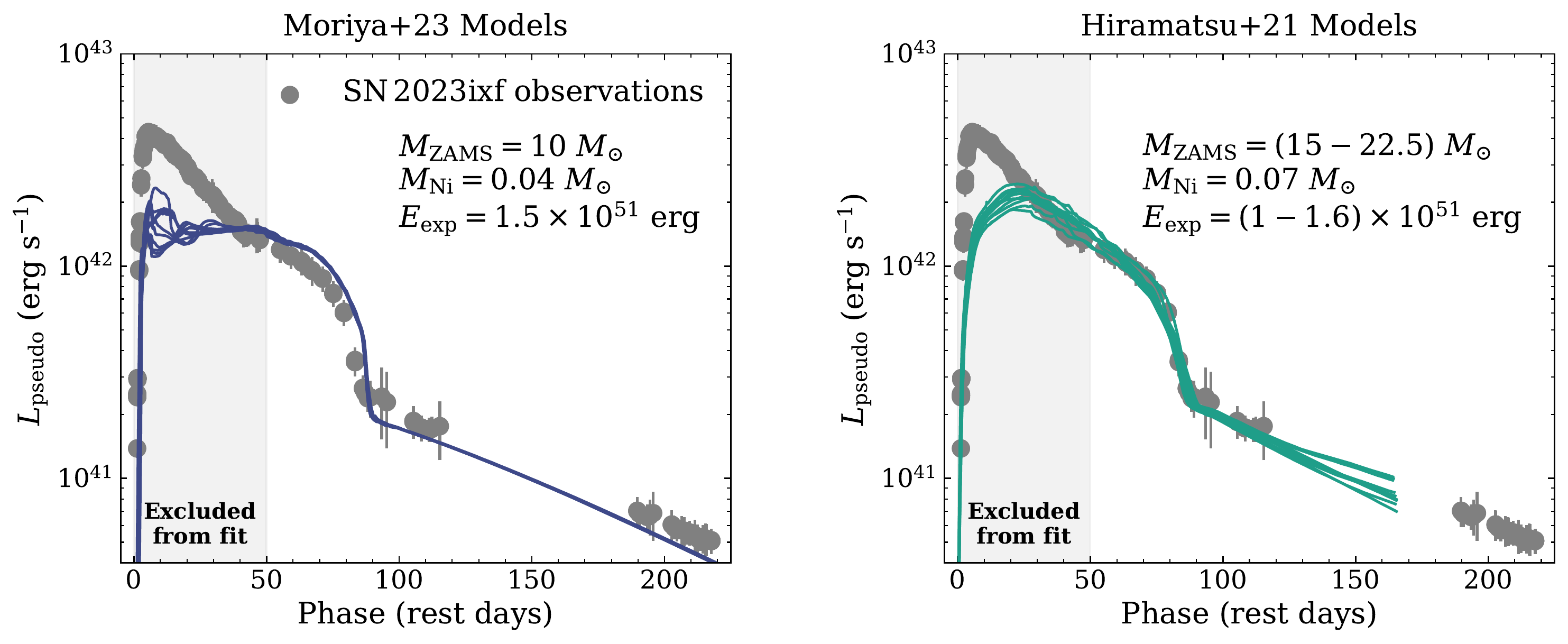}
    \caption{Comparison of our bolometric light curve to models from \cite{Moriya_2023} (left) and \cite{Hiramatsu_2021} (right). The regions shaded in light gray indicate epochs excluded from the $\chi^2$ fitting procedure. Here, $M_{\rm final}$ refers to the mass of each corresponding progenitor model at core collapse. The models from \cite{Moriya_2023} match the plateau duration, but systematically overestimate the plateau luminosity and underestimate the tail luminosity. On the other hand, the models from \cite{Hiramatsu_2021} reproduce the plateau and nebular-phase behaviors of SN~2023ixf well.}
    \label{fig:model_grids}
\end{figure*}

\subsection{Moriya et al.\ Accelerated Wind Models}
\label{sec:moriya_models}

We obtain the large grid of pre-computed multi-frequency hydrodynamic models from \cite{Moriya_2023}. The grid\footnote{Accessible at \url{https://datadryad.org/dataset/doi:10.5061/dryad.pnvx0k6sj} (\citealt{Moriya_grid}).} contains 228,016 synthetic light curves simulated using the multi-group radiation hydrodynamics code {\tt STELLA} (\citealt{Blinnikov_1998,Blinnikov_2000,Blinnikov_2006,Blinnikov_2004,Baklanov_2005}), based on five RSG progenitor models with initial masses of 10, 12, 14, 16, and 18 $M_\odot$ from \cite{Sukhbold_2016}, evolved with {\tt KEPLER} (\citealt{Weaver_1978}). Confined CSM density structures following a $\beta$-law wind velocity profile are attached on top the progenitor models, with varying mass-loss rates, CSM extents, and wind acceleration parameters. The models span a wide range of explosion and CSM parameters, with $E_{\rm exp}=(0.5-5)\times10^{51}\ {\rm erg}$, $M_{\rm Ni}=(0.001-0.3)\ M_{\odot}$, $\dot{M}=(10^{-5}-10^{-1})\ M_{\odot}\ {\rm yr^{-1}}$, $R_{\rm CSM}=(1-10)\times10^{14}\ {\rm cm}$, and $\beta=0.5-5.0$. More details regarding the setup of the model grid can be found in \cite{Moriya_2023}. We present a few of the top models selected via $\chi^2$ minimization in the left panel of Figure~\ref{fig:model_grids}. 

The top models ($\chi^2\approx 4800$) from the \cite{Moriya_2023} model grid all exhibit the same behavior of overestimating the plateau luminosity and underestimating the tail luminosity, but replicate the plateau duration well. They all share the same progenitor and explosion parameters: an initial mass of $M_{\rm ZAMS}=10\ M_{\odot}$ (or a final mass of $M_{\rm final}=9.7\ M_{\odot}$), a pre-explosion radius of $R=510\ R_{\odot}$, an explosion energy of $E_{\rm exp}=2\times10^{51}$ erg, and a nickel mass of $M_{\rm Ni}=0.06\ M_{\odot}$. Qualitatively, these models behave similarly to those recovered in \cite{Moriya_2024}, with the main difference being the derived $M_{\rm Ni}$, which is likely the result of fitting the bolometric light curve in this work as opposed to fitting the pseudo-bolometric light curve constructed with only optical coverage.

We additionally perform the same $\chi^2$ fits to the plateau phase ($30-90$ days) and the early nebular phase ($90-150$ days) independently (not shown). We find that models with $M_{\rm ZAMS}=12\ M_{\odot}$, $E_{\rm exp}=1.5\times10^{51}\ {\rm erg}$, and $M_{\rm Ni}<0.01\ M_{\odot}$ can replicate the plateau behaviors, while models with $M_{\rm ZAMS}=12\ M_{\odot}$, $E_{\rm exp}=5\times10^{51}\ {\rm erg}$, and $M_{\rm Ni}=0.08\ M_{\odot}$ can replicate the early-nebular-phase behaviors. When these parameter values are combined, however, we either find inflated plateau luminosity or underestimated tail luminosity, with the plateau duration displaced by up to 10 days. Even if we restrict the timescale to $50-120$ days to minimize any potential effects of obscured CSM interaction during the plateau and avoid phases on the radioactive tail where the assumptions of hydrodynamic models begin to break down, we still find that no meaningful combination of $M_{\rm ZAMS}$, $E_{\rm exp}$, and $M_{\rm Ni}$ in the model grid can reproduce properties on the plateau and the radioactive tail simultaneously.

There may be a combination of several reasons for why the model grid from \citet{Moriya_2023} is unable to reproduce the plateau and nebular-phase luminosity of SN~2023ixf. Most important of all is perhaps the established relationship between plateau properties and progenitor structures, such as the radius and the hydrogen-rich envelope mass (e.g., \citealt{Popov_1993, Kasen_Woosley_2009, Dessart_2013,Moriya_2016}). In this vein, the stellar models from \cite{Sukhbold_2016} may not reflect the true properties of SN~2023ixf's progenitor. In addition, the plateau properties also depend critically on the explosion energy and synthesized nickel mass, as they collectively set the extent of plateau luminosity, which affects the subsequent evolution of H-recombination front (see Section~\ref{sec:degeneracy}). Moreover, the non-uniqueness of light-curve properties as a function of progenitor mass in SNe IIP has been discussed in detail (e.g., \citealt{Dessart_2019, Goldberg_2019, Goldberg_2020}), especially when considering variations in stellar structure due to varied stellar mass loss (e.g., \citealt{Morozova_2015, Hiramatsu_2021, Dessart_2024}). 

\subsection{Hiramatsu et al.\ CSM-Free Models}
\label{sec:hiramatsu_models}

Motivated by reasons outlined in Section~\ref{sec:moriya_models}, we retrieve the CSM-free model grids from \cite{Hiramatsu_2021} constructed using Modules for Experiments in Stellar Astrophysics ({\tt MESA}; \citealt{Paxton_2011,Paxton_2013,Paxton_2015,Paxton_2018,Paxton_2019,Jermyn_2023}) + \texttt{STELLA}, and compare them with our observations in the right panel of Figure~\ref{fig:model_grids}. The advantages of the models from \cite{Hiramatsu_2021} are that they survey a wider range of progenitor mass ($10-25\ M_\odot$ in increments of $2.5\ M_\odot$) and take various mass-loss efficiencies ($\eta_{\rm wind}=0-3$ in increments of 0.1) into consideration over the lifetime of each RSG progenitor. These extra measures result in a more diverse array of progenitor structures prior to explosion compared to those five unique progenitor masses from \cite{Sukhbold_2016} used in \cite{Moriya_2023}. The caveats of these models are that the range of $E_{\rm exp}=(0.5-2.0)\times10^{51}\ {\rm erg}$ (in increments of $0.5\times10^{51}\ {\rm erg}$) and $M_{\rm Ni}=0.04,0.07,0.1\ M_{\odot}$ employed are much more restrictive than the models from \cite{Moriya_2023}.

After performing the same $\chi^2$ fitting procedure to the bolometric light curve from $30-150$ days after explosion, we find that lower-mass stars that undergo moderate mass loss and higher-mass stars that undergo intense mass loss can equally reproduce the light-curve behaviors of SN~2023ixf ($\chi^2\approx300-500)$. Models with high-mass progenitors ($M_{\rm ZAMS}\gtrsim 15\ M_{\odot}$) lose $>30\%$ of their initial mass, which may be unrealistically high for single RSGs (e.g., \citealt{Beasor_2020}) undergoing standard stellar evolution, and we will discuss the implications of this later in the paper. Most of these progenitor models also have a larger pre-explosion radius ($R\approx460-820\ R_{\odot}$) and a lower envelope mass ($M_{\rm H,env}\approx 3-5\ M_{\odot}$) compared to the best-fit models from Section~\ref{sec:moriya_models}. We derive a $^{56}$Ni mass of $0.07\ M_{\odot}$ here, which is consistent with $M_{\rm Ni}=0.068\ M_{\odot}$ from Section~\ref{sec:plateau_fit}. The explosion energies of $E_{\rm exp}=1.2-1.8\times10^{51}\ {\rm erg}$ recovered here are also systematically lower than $E_{\rm exp}=2\times10^{51}\ {\rm erg}$ found in Section~\ref{sec:moriya_models}, which correspond to models with lower $M_{\rm final}$ that require lower energies to match observed properties for a given radius (see Section~\ref{sec:degeneracy}).

Given that relatively low-mass progenitors with moderate mass loss and high-mass progenitors with enhanced mass loss can both reproduce plateau behaviors, additional constraints must be imposed to further discern the physical origin of SN~2023ixf.

\section{Constraining Progenitor and Light-Curve Models with Pulsation Period}
\label{sec:rsg_mesa_gyre}

\subsection{Degeneracy in Type IIP SN Light Curves}
\label{sec:degeneracy}

An important caveat of the model grid from \cite{Moriya_2023} is the fixed pre-explosion mass-radius relationship of the progenitors. In fact, this caveat applies to any attempt to derive a pre-explosion mass from a grid of progenitor models where each underlying progenitor star exploded is identified with a single progenitor radius at the time of explosion. As shown in \cite{Goldberg_2019} and \cite{Goldberg_2020} (see also discussions in \citealt{Martinez_Bersten_2019} and \citealt{Dessart_2019}), there exist degeneracies (``a family of explosions") between progenitor radius, ejecta mass, and explosion energy that cannot be lifted with one-to-one mappings of pre-explosion progenitor properties and explosion properties. These degeneracies can be well-described by the scaling relations of \cite{Popov_1993}, \cite{Kasen_Woosley_2009}, \cite{Sukhbold_2016}, \cite{Goldberg_2019}, and others.

\begin{figure}[t!]
    \centering
    \includegraphics[width=\columnwidth]{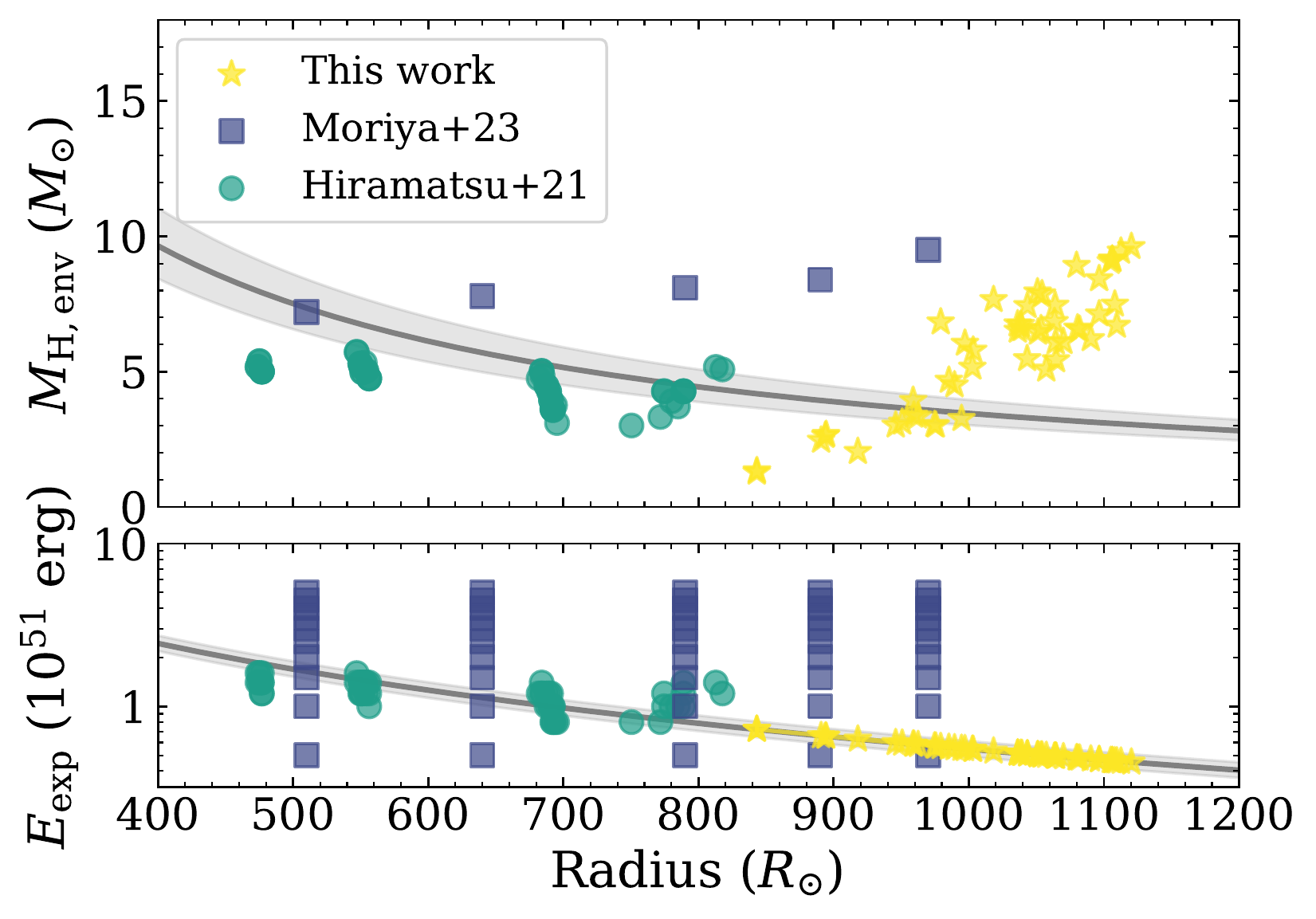}
    \caption{Degeneracy curves for SN~2023ixf recovered from the scaling relations of \cite{Goldberg_2019} as a function of progenitor radius. The gray solid lines indicate the H-rich envelope mass (top panel) and explosion energy (bottom panel) required for a given progenitor radius to match the observed SN properties such as plateau duration and luminosity at day 50. For demonstration purposes, the degeneracy curves were constructed with a nickel mass of $M_{\rm Ni}=0.068\ M_{\odot}$. Shaded regions correspond to $1\sigma$ bounds that take the intrinsic scatter of the scaling relations and errors on the parameters used into account. We also show models from \cite{Moriya_2023} (navy squares; all possible combination), \cite{Hiramatsu_2021} (green circles; top 50 models selected via $\chi^2$ fitting), and our own modeling (yellow stars; models selected via pulsation period).} 
    \label{fig:degeneracy}
\end{figure}

We show the corresponding degeneracy curves for SN~2023ixf using the scaling relations of \cite{Goldberg_2019} in Figure~\ref{fig:degeneracy} (their Eq.~22), calculated with the bolometric luminosity at day 50 ($L_{50}$), plateau duration ($t_p$), and $^{56}$Ni mass ($M_{\rm Ni}$), overplotted with the models from \cite{Moriya_2023} and \cite{Hiramatsu_2021}. Instead of plotting $M_{\rm ej}$, we show the H-rich envelope mass $M_{\rm H, env}$ in Figure~\ref{fig:degeneracy}. The original argument for using $M_{\rm ej}$ to calibrate the scaling relations in \cite{Goldberg_2019} was that their models exhibited strong mixing of hydrogen deep into the interior of the star from self-consistent mixing via the \cite{Duffell_2016} Rayleigh-Taylor Instability (RTI) mixing prescription. This resulted in the majority of the ejecta partaking in hydrogen recombination and thus collectively driving the evolution of the SN on the plateau. However, the models in \cite{Moriya_2023} (with progenitors from \citealt{Sukhbold_2016}), \cite{Hiramatsu_2021}, and our own (see the following subsections) contain progenitors that only have little to no hydrogen mixed into the core. Additionally, in partially-stripped envelopes, the H-rich envelope mass makes up a smaller fraction of the total ejecta mass. Therefore, the H envelope mass at the time of explosion is the more appropriate choice to use for comparing to scaling relations here, as it drives the bulk evolution of the luminosity during the photospheric phase. 

From Figure~\ref{fig:degeneracy}, it is not surprising to see that the best-fitting models from the pre-computed grid of \cite{Moriya_2023} are all consistent with a low-mass progenitor with a higher explosion energy, as the $10\ M_{\odot}$ (with $M_{\rm H,env}=7.2\ M_{\odot}$) progenitor model exploded with $E_{\rm exp}=(1.5-2)\times10^{51}$ erg perfectly intersect both degeneracy curves. The same argument goes for the initially high-mass progenitors from \cite{Hiramatsu_2021} that lose significant amounts of mass and end up with larger radii prior to core collapse, thereby only requiring $E_{\rm exp}\approx 10^{51}$ erg (note as well that these models all have similar remaining H envelope mass of $\approx 4-7\ M_{\odot}$, due to the stronger mass-loss prescription adopted in more massive stars). Models that intersect the degeneracy curves are also the best-fitting models selected via $\chi^2$ distributions in Section~\ref{sec:model_desc}, further supporting the choice of using $M_{\rm H,env}$ instead of $M_{\rm ej}$. The ability of these simple scaling laws in predicting explosion properties highlights the crucial need for independent measurements of the progenitor radius in aiding hydrodynamical modeling of SN IIP light curves. 

Another important measurement that may provide an alternative constraint on progenitor and explosion properties is the expansion velocity, or the photospheric velocity, as it encodes information regarding progenitor and explosion properties (i.e., $M_{\rm ej}$, $R$, and $E_{\rm exp}$). The expansion velocity is typically inferred from the absorption minimum of the \ion{Fe}{2} $\lambda$ 5169 line (\citealt{Dessart_Hillier_2005,Pejcha_2015,Gutierrez_2017,Muller_2017}). 
Several works in the literature (e.g., \citealt{Martinez_Bersten_2019,Ricks_2019,Martinez_2020,Martinez_2022b}) argue that simultaneously fitting the bolometric light curve and photospheric velocity evolution can help estimate the physical properties of SNe II given the significant dispersion and outliers on the observed $L-v_{\rm ph}$ relationship (e.g., \citealt{Pejcha_2015, Gutierrez_2017,Rodriguez_2020}). There are also several studies establishing the existence of a ``standard candle" relation between $L$ and $v_{\rm ph}$, which was first discovered observationally by \cite{Hamuy_2003} and explained physically by \cite{Kasen_Woosley_2009}. This $L\propto v_{\rm ph}^2$ relation (up to a dilution factor that is taken to be a function only of the luminosity; \citealt{Wagoner_1981,Eastman_1996}) during the photospheric phase serves as the foundation for the expanding photosphere method for measuring distances to SNe (\citealt{Kirshner_Kwon_1974}), as well as the standardized candle method for SNe IIP cosmology (e.g., \citealt{Maguire_2010,Olivares_2010,deJaeger_2020,Vogl_2024}). In this context, other works (e.g., \citealt{Goldberg_2019,Goldberg_2020, Fang_2025}) have argued against using $v_{\rm ph}$ during the plateau phase as an independent parameter for breaking light-curve degeneracies. Given the significant divide surrounding this topic, we focus our discussion only on light-curve degeneracies on the plateau luminosity.

Due to the high circumstellar extinction of SN~2023ixf's progenitor (\citealt{Jencson_2023,Kilpatrick_2023,Soraisam_2023,Xiang_2024,Van_Dyk_2024}), the measurement of pre-explosion radius---which may help identify the explosion parameters---is a non-trivial task accompanied by significant uncertainty. In the next section, we seek to remedy our inability to accurately determine the progenitor radius from pre-explosion imaging by creating an additional small grid of {\tt MESA} models with a wide range of progenitor properties, and narrow down viable candidate models via observed variability. We stress here that the mass derived from hydrodynamic modeling reflects the pre-explosion state of the progenitor, and any inference regarding the ZAMS conditions of the progenitor depends sensitively on the physical assumptions (such as mass loss and convective efficiency) adopted in the stellar evolution code.

\subsection{{\tt MESA} Progenitor Models Matching Observed Progenitor Variability}
\label{sec:mesa_models}

An important, unique constraint on the nature of SN~2023ixf's progenitor comes from the observed IR variability with a perodicity of $P\approx 1100$ days \citep{Soraisam_2023, Jencson_2023, Kilpatrick_2023}. RSG variability has been well-studied in galactic and nearby stellar populations (e.g., \citealt{Jurcevic_2000, Kiss_2006, Percy_2014, Soraisam_2018, Conroy_2018, Chatys_2019, Ren_2019}), with periods typically ranging from a few hundred to a few thousand days with period-luminosity relations characteristic of fundamental-mode and first-overtone radial pulsations. While the pulsation phase may impact the light curves via the $\approx$10\% variation in the progenitor radius at the time of explosion \citep{Goldberg_2020a}, RSG pulsation periods are mostly sensitive to the star's luminosity and envelope density \citep{Stothers_1969,Guo_2002,Joyce_2020}.
We thus construct additional models in order to incorporate additional information about the stellar structure provided by the observed progenitor variability.

Using {\tt MESA} revision 23.05.1, we construct non-rotating, solar-metallicity ($Z=0.02$)\footnote{\cite{Van_Dyk_2024} found that the metallicity at the SN site is compatible with subsolar to supersolar values, but for simplicity here, we only consider solar metallicity.} models with varying ZAMS masses ($M_{\rm ZAMS}=10–25\ M_{\odot}$ in increments of $0.5\ M_{\odot}$) and convective efficiencies ($\alpha_{\rm MLT}=1.5–3$ in increments of 0.5) in the hydrogen-rich envelope. We note that varying $\alpha_{\rm MLT}$ provides additional variation in the stellar radius \citep{Stothers_1995,Massey_2003,Meynet_2015,Goldberg_2022a}. We adopt {\tt MESA}’s “Dutch” prescription for winds with varying efficiency factor ($\eta_{\rm wind}=0–3$ in increments of 0.5), and use modest convective overshooting parameters $f_{\rm ov} = 0.01$ and $f_{\rm 0,ov} = 0.005$. For reference, a wind efficiency of $\eta_{\rm wind}=1$ under the Dutch scheme translates to a mass-loss rate that is $\sim10$ higher than measured values for normal RSG winds ($\dot{M}_{\rm wind}\lesssim 10^{-5}\ M_{\odot}\ {\rm yr^{-1}}$; e.g., \citealt{Beasor_2020}). Other inputs are determined following the {\tt 12M\_pre\_ms\_to\_core\_collapse} case of the {\tt MESA} test suite, which is based on the setup described in Section~6 of \citet{Paxton_2018} and \citet{Farmer_2016}. 

\begin{figure}[t!]
    \centering
    \includegraphics[width=\columnwidth]{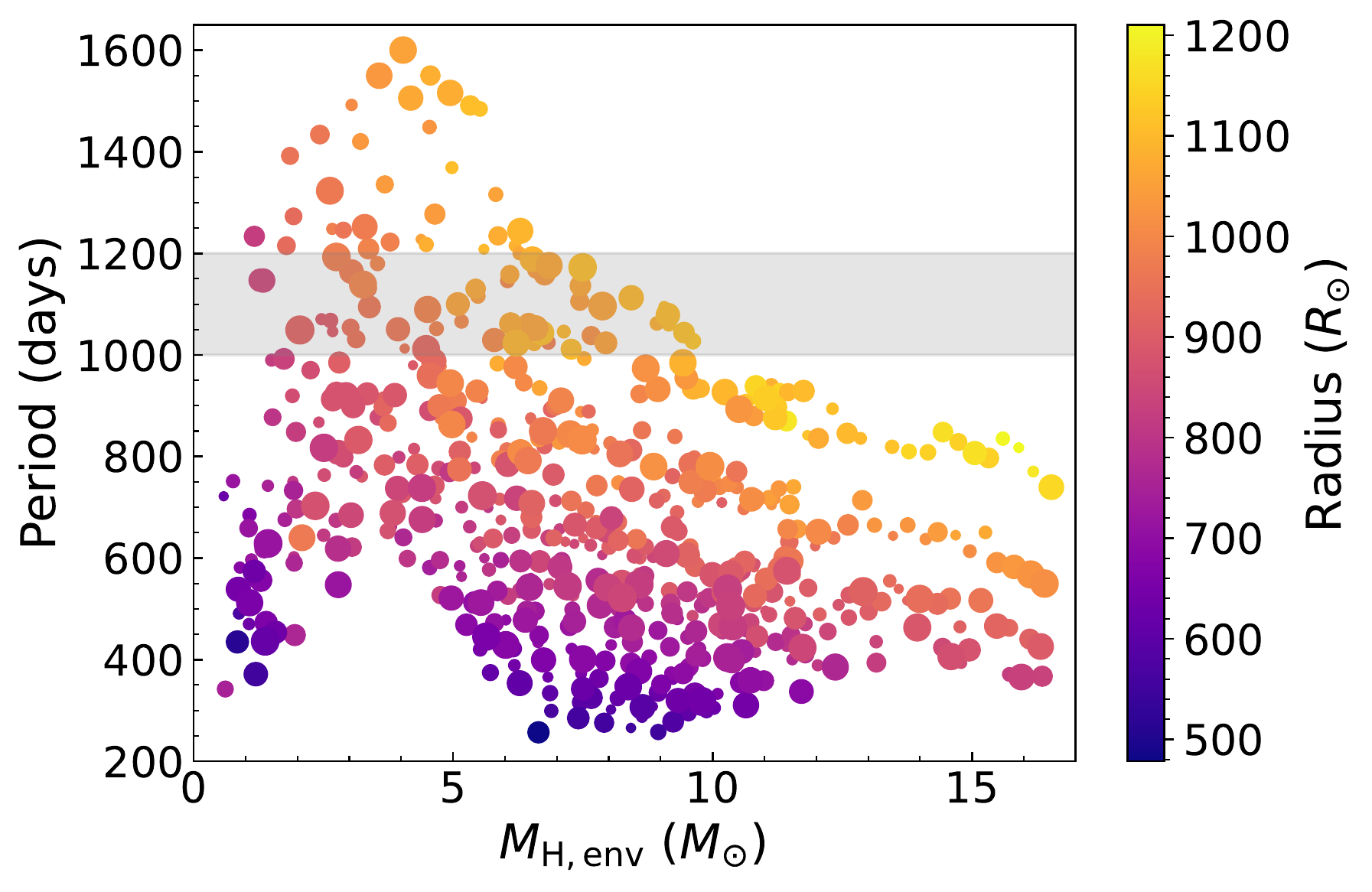}
    \caption{Pulsation period at $\approx$7000 days before core collapse as a function of the hydrogen envelope mass $M_{\rm H,env}$ at explosion. Each model is color-coded by the progenitor radius and the size corresponds to the initial mass $M_{\rm ZAMS}$. The shaded region in gray highlights the models with periods within the range $P=1000-1200$ days, consistent with the observed IR variability in \cite{Soraisam_2023} and \cite{Jencson_2023}. Models that are consistent with the observed variability have $M_{\rm ZAMS}=13-24\ M_{\odot}$ and final masses of $M_{\rm final}\approx6.5-15.46\ M_{\odot}$ (which correspond to envelope masses of $M_{\rm H, env}\approx1.28-9.26\ M_{\odot}$) at the time of core collapse.}
    \label{fig:periods}
\end{figure}

To compare the candidate RSG progenitor of SN~2023ixf, which exhibited strong variability with a long baseline period of $P\approx 1100$ days (\citealt{Jencson_2023,Soraisam_2023}), we use the the pulsation instrument {\tt GYRE} version 7.2 (\citealt{Townsend_Teitler_2013}) to identify the period for the fundamental radial ($l=0$) mode at $\approx7000$ days prior to the onset of core collapse (which loosely translates to core C burning). {\tt GYRE} is based on a Magnus Multiple Shooting scheme and provides both adiabatic and non-adiabatic solutions to the linearized pulsation equations for high-resolution structural models produced during the RSG phase. We consider only adiabatic results in this analysis. We plot the period as a function of the hydrogen envelope mass $M_{\rm H,env}$ and progenitor radius at core collapse in Figure~\ref{fig:periods}, which shows that a low envelope mass ($\lesssim 10\ M_{\odot}$) and a large progenitor radius ($\gtrsim 800\ R_{\odot}$) are required to match the observed variability. While model uncertainties on pulsation periods may have an effect on our analysis here, they are largely restricted to resolution effects (\citealt{Li_2025}) at well under $\sim 10\%$ for RSG variability, and should be captured by the large range of $\alpha_{\rm MLT}$ considered here (\citealt{Joyce_2020}). In any case, we do not expect any model uncertainties to exceed or approach the uncertainty of the observed pulsation period.

Another caveat of our model grid, similar to the findings in \cite{Hiramatsu_2021}, is that many low-mass models ($M_{\rm ZAMS}\lesssim 12.5M_{\odot}$) develop degenerate cores and fail to converge during later burning stages that require off-center ignition (O, Ne, or Si). While excluding these models may introduce some bias to our sample, most of these low-mass progenitors likely will not have the correct physical properties (e.g., a low envelope mass with a larger radius; see Figure~\ref{fig:periods}) required to satisfy the high fundamental-mode pulsation period observed for SN~2023ixf. For reference, the highest pulsation period at the end of core C burning we find for these models is $P\approx 975$ days ($M_{\rm ZAMS}=12.5\ M_{\odot}$, $\eta_{\rm wind}=2.0$, and $\alpha_{\rm MLT}=1.5$), outside the selection criteria of $P=1000-1200$ days we impose on period. We therefore discard these models at this stage.

We continue evolving models with pulsation period in the range of $P=1000-1200$ days through core collapse, then explode the stellar models via a thermal bomb energy deposition by modifying the {\tt ccsn\_IIp} test suite before handing off to the radiation-hydrodynamics instrument {\tt STELLA} to create synthetic observables. While exploding progenitor models at the appropriate pulsational phase by injecting a velocity profile proportional to the fundamental mode eigenfunction as done in \cite{Goldberg_2020a} can provide additional constraints, it is outside the scope of this paper and we leave it as an avenue for improved modeling in the future. As first-order estimates for progenitor and explosion properties, we return to the scaling relations to calculate the explosion energy $E_{\rm exp}$ required to match the plateau properties. Since the scaling relations, as well as $L_{50}$ and $t_p$, have intrinsic uncertainties associated with them, we also capture the $1\sigma$ bounds for $E_{\rm exp}$ for a total of three explosion energies for each progenitor model and spanning a total range of $E_{\rm exp}=(0.24-0.87)\times10^{51}\ {\rm erg}$. Due to the large progenitor radii and low envelope masses required to reproduce the observed pulsation period, most of our progenitor models favor relatively low-energy explosions ($E_{\rm exp}<10^{51}\ {\rm erg}$) in order to simultaneously reproduce the light curve brightness and duration, contrary to previous findings (\citealt{Bersten_2024,Moriya_2024,Singh_2024}). The SN shock propagation is modeled following the prescription of RTI mixing \citep{Duffell_2016} until near shock breakout, with the final $^{56}$Ni scaled to match a total mass of $M_{\rm Ni}=0.068M_{\odot}$ calculated in Section~\ref{sec:plateau_fit}. Here, we do not perform any additional mixing (e.g., boxcar) of $^{56}$Ni into the hydrogen envelope other than the abundance distribution given by RTI mixing, as it is not clear if and how far $^{56}$Ni is mixed beyond the core (see, e.g.,  \citealt{Kasen_Woosley_2009,Bersten_2011,Dessart_2013,Moriya_2016}, for the effects of $^{56}$Ni mixing on the characteristics of the decay tail).

\begin{deluxetable*}{lccccccccccc}
\tablewidth{\textwidth} 
\caption{Progenitor and Explosion Properties of our {\tt MESA+STELLA} Models}
\tablehead{\colhead{} & 
\colhead{$M_{\rm ZAMS}$} & 
\colhead{$\eta_{\rm wind}$} & 
\colhead{$\alpha_{\rm MLT}$} &
\colhead{$M_{\rm final}$} &
\colhead{$M_{\rm H,env}$} &
\colhead{$R$} &
\colhead{$P$} &
\colhead{$E_{\rm exp}$} &
\colhead{$M_{\rm ej}$} &
\colhead{$M_{\rm fallback}$} &
\colhead{$\chi^2$}\\[-12pt]
\colhead{Progenitor Model} & 
\colhead{} & 
\colhead{} & 
\colhead{} &
\colhead{} &
\colhead{} &
\colhead{} &
\colhead{} &
\colhead{} &
\colhead{} &
\colhead{} &
\colhead{} \\[-12pt]
\colhead{} & 
\colhead{($M_{\odot}$)} & 
\colhead{} & 
\colhead{} &
\colhead{($M_{\odot}$)} &
\colhead{($M_{\odot}$)} &
\colhead{($R_{\odot}$)} &
\colhead{(days)} &
\colhead{($10^{51}$ erg)} &
\colhead{($M_{\odot}$)} &
\colhead{($M_{\odot}$)} &
\colhead{}
}
\label{tab:mesa_models}
\startdata
16.5M\_eta2.5\_alpha2.0 & 16.5 & 2.5 & 2.0 & 8.53 & 3.39 & 959 & 1094 & 0.70 & 6.48 & 0.40 & \phn297.10\\
17.5M\_eta2.5\_alpha2.0 & 17.5 & 2.5 & 2.0 & 8.74 & 3.26 & 995 & 1139 & 0.66 & 6.21 & 0.78 & \phn318.37\\
18.0M\_eta2.5\_alpha1.5 & 18.0 & 2.5 & 1.5 & 8.53 & 3.02 & 976 & 1167 & 0.68 & 6.05 & 0.80 & \phn597.14\\
21.0M\_eta1.5\_alpha1.5 & 21.0 & 1.5 & 1.5 & 9.95 & 3.35 & 962 & 1131 & 0.70 & 6.83 & 1.25 & \phn657.83\\
18.0M\_eta1.5\_alpha1.5 & 18.0 & 1.5 & 1.5 & 8.80 & 3.04 & 975 & 1164 & 0.68 & 5.79 & 1.18 & 1050.54\\[+2pt]
\enddata
\tablecomments{Here, $M_{\rm final}$, $M_{\rm H,env}$, and $R$, are the final stellar mass, hydrogen-rich envelope mass, and radius at the time of core collapse, $P$ is the fundamental-mode pulsation period at $\approx 7,000$ days before core collapse, and $M_{\rm ej}$ and $M_{\rm fallball}$ are the surviving ejecta mass and fallback mass at the time of shock breakout.}
\end{deluxetable*}

As this work is concerned with the energetics of SN~2023ixf derived from the plateau, we only explore CSM-free explosion models. We use 800 spatial zones for the SN ejecta and 100 frequency bins to yield convergence in the synthetic bolometric light curves produced by {\tt STELLA}. Any late-time fallback material produced by reverse shocks from core boundaries interacting with the lowest-velocity inner material is excised via the zero-energy technique described in \cite{Paxton_2019} and \cite{Goldberg_2019}, with an additional standard velocity cut of 500 km s$^{-1}$ at the inner boundary at handoff between {\tt MESA} and {\tt STELLA}. We find that the low explosion energies inferred for our models are not significantly larger than the magnitude of the binding energy of their progenitors at the time of explosion. As such, we observe non-negligible fallback, with $M_{\rm fallback}\approx 0.5-3.6\ M_{\odot}$. This is also seen in models of some other short-plateau SNe IIP (e.g., \citealt{Teja_2024}). These materials that do not collapse into the initial compact remnant object could supply additional accretion-powered luminosity that interacts with the ejecta and thereby alter the observed light-curve properties (e.g., \citealt{Dexter_2013, Chan_2018, Lisakov_2018,Moriya_2019}). We also note that only the innermost ejecta layers experience fallback, so plateau properties mediated by H-recombination might be unaffected until the fall from the plateau, with other signatures of fallback accretion possibly manifesting at much later times. Since the proper treatment of late-time fallback and subsequent accretion onto the proto-neutron star in 1D simulations remains an open question, we therefore caution that our models should be viewed as approximations and may not reflect all physical processes at work. 

The {\tt MESA+GYRE+STELLA} inlists used to generate the models in this work are available on Zenodo under an open-source Creative Commons Attribution 4.0 International license: \dataset[10.5281/zenodo.15368365]{https://zenodo.org/records/15368365}.

\subsection{Comparison to Observations}
\label{sec:comp_mods}

\begin{figure*}[t!]
    \centering
    \includegraphics[width=\textwidth]{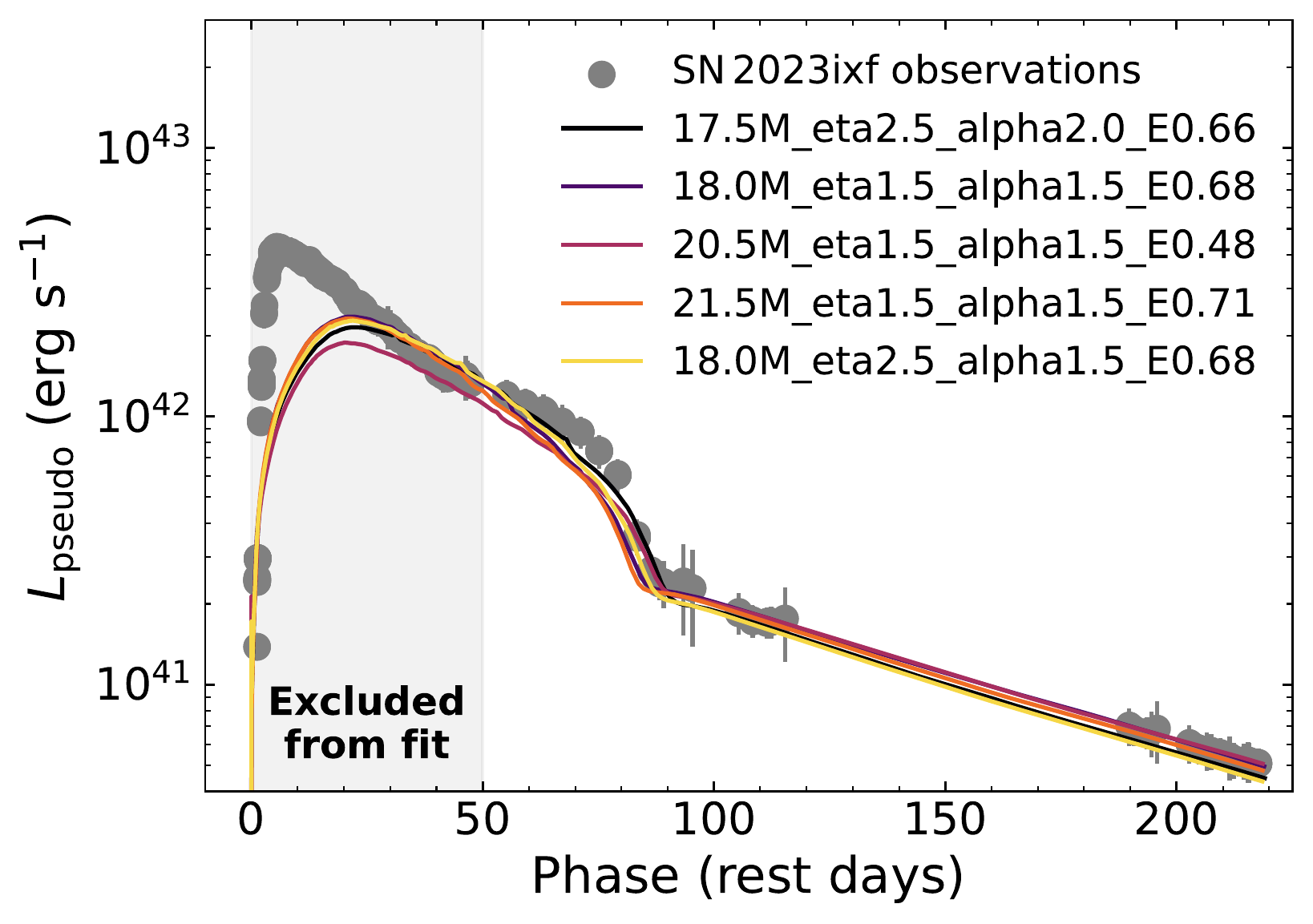}
    \caption{Top left: Bolometric light curves of the top five best-fit CSM-free models for SN~2023ixf. The region shaded in light gray indicates epochs excluded from the $\chi^2$ fitting procedure. With the additional constraints from pulsation period and the degeneracy curves, our synthetic light curves recover the plateau properties well. The measured value of $M_{\rm Ni}=0.068\ M_{\odot}$ provides an excellent agreement with observations on the radioactive tail. Bottom left: The photospheric and \ion{Fe}{2} $\lambda$ 5169 velocity estimates of our best-fit models compared to the observed evolution of \ion{Fe}{2} $\lambda$ 5169 line velocity from \cite{Singh_2024}. Models that match the plateau luminosity produce nearly identical velocity evolution during the plateau phase. Right: The corresponding optical multi-band light curves of the best-fit models.}
    \label{fig:no_csm}
\end{figure*}

Following the procedure in Section~\ref{sec:model_desc}, we calculate the $\chi^2$ values of our {\tt MESA+STELLA} models. We list the progenitor and explosion properties of our top five best-fit models ($\chi^2\approx300-1000$) in Table~\ref{tab:mesa_models} and show the corresponding synthetic bolometric light curves in the top left panel of Figure~\ref{fig:no_csm}. The naming convention follows $\langle M_{\rm ZAMS}\rangle$M\_eta$\langle\eta_{\rm wind}\rangle$\_alpha$\langle\alpha_{\rm MLT}\rangle$\_E$\langle E_{\rm exp}\rangle$. With a more detailed modeling that follows the degenerate scaling relations, we recover higher-mass progenitors ($M_{\rm ZAMS}\gtrsim16.5\ M_{\odot}$) that shed a significant portion of their hydrogen envelopes to match the plateau behaviors best, similar to our findings in Section~\ref{sec:hiramatsu_models}. These models are in good agreement with observations, albeit slightly underluminous during later plateau phases ($t\gtrsim 60$ days). The plateau duration and the radioactive tail, however, are in excellent agreement with observations. 

In the bottom left panel of Figure~\ref{fig:no_csm}, we compare the expansion velocity of our models to the observed \ion{Fe}{2} $\lambda$ 5169 velocity evolution \citep{Singh_2024}, estimated from the blueshifted absorption trough of the line profile. The line velocity has a systematic uncertainty of $\sim500\ {\rm km\ s^{-1}}$ dominated by the instrumental resolution ($\sim10$ \AA). We show both the photospheric velocity and the \ion{Fe}{2} velocity from our best-fit models. The photospheric velocity is calculated at a Rosseland optical depth of $\tau_{\rm Ros}=2/3$ and the \ion{Fe}{2} line velocity is approximated in the Sobolev approximation at a Sobolev optical depth of $\tau_{\rm Sob}=1$. Note that we are not simultaneously fitting both the bolometric light curve and the observed velocity here; we fit only the luminosity and recover agreement in model velocities. 

The photospheric velocities of our best-fit models agree with each other, and give satisfactory matches with observations up to $\sim 45$ days before diverging. This is because the photosphere migrates inward in mass coordinate as the ejecta expand, and the \ion{Fe}{2} $\lambda$ 5169 line is expected to form exterior to the photosphere (see a detailed discussion in \citealt{Paxton_2018}). At later times, the divergence between the line-forming region and the photosphere becomes more pronounced, as the decay of $^{56}$Ni and $^{56}$Co increases the overall Fe abundance and slows the inward progression of the line-forming region in mass coordinate (\citealt{Sobolev_1960,Castor_1970,Mihalas_1978,Kasen_2006}). Attempts to account for this deviation by raising the explosion energy would cause a mismatch to the luminosity and overestimate the early-time velocity. Although, the same models are also underluminous in bluer bands (see right panel in Figure~\ref{fig:no_csm}), and accounting for the color information may help alleviate the discrepancy in photospheric velocity during the late plateau phase. Contrarily, the model \ion{Fe}{2} velocities approximated at $\tau_\mathrm{Sob}=1$ are systematically higher than observations. This is not concerning, as the precise Sobolev optical depth where the \ion{Fe}{2} line is formed is not known, and evaluating the velocity at a higher $\tau_{\rm Sob}$ closer to the location of the photosphere can resolve this discrepancy, as might geometric effects due to deviations from spherical symmetry (\citealt{Vartanyan_2025}). Likewise, a lower metallicity that is consistent with the environmental study of SN~2023ixf presented in \cite{Van_Dyk_2024} or an asymmetric explosion (\citealt{Fang_2024,Li_2024}) can also explain the faster model \ion{Fe}{2} velocities. While these models show satisfactory qualitative agreement with the observed velocities, none of our models, including those not shown which provide weaker fits to the bolometric light curve, show exceptional agreement with entire observed velocity evolution. This highlights the complex physics at play that govern the evolution of SN~2023ixf and the nature of its progenitor and environment.

In the right panel of Figure~\ref{fig:no_csm} we show the corresponding optical multi-band light curves for the same models, which are constructed by convolving the SEDs returned by {\tt STELLA} with the transmission function of each filter ($UBVgriz_s$) at each timestep. The luminosity excess in both the observed bolometric light curve and multi-band light curves, as well as the higher observed photospheric velocity compared to our models at $t\gtrsim 60$ days, could be explained by the presence of a less dense and extended CSM structure (e.g., \citealt{Morozova_2017, Singh_2024}), which supplies additional energy. In the case of SN~2023ixf, sustained CSM interaction during the plateau phase has been confirmed both photometrically and spectroscopically, with estimated densities of $\sim10^{-20}-10^{-16}\ {\rm g\ cm^{-3}}$ (or mass-loss rates of $10^{-5}-10^{-3}\ M_{\odot}\ {\rm yr^{-1}}$ if we assume a steady-state wind) for the material ejected from its RSG progenitor $\approx7-33$ years prior to explosion (\citealt{Bostroem_2024,Singh_2024}). It is worth noting that even with this additional CSM interaction, the ejecta is not expected to slow down significantly (\citealt{Dessart_2017}). The excess on the radioactive tail in redder bands could be attributed to radiation leaking out from bluer bands, and a more centrally concentrated $^{56}$Ni distribution could potentially resolve the discrepancies in colors.

\section{Discussion}
\label{sec:discussion}

\subsection{Differences Compared to Previous Work}
\label{sec:previous_work}

SN~2023ixf has been the object of several investigations since its discovery, ranging from pre-explosion observations of the SN site to properties of the SN itself. The wealth of pre-explosion observations and detailed studies of the early evolution provide constraints on the progenitor that are among the best of any modern SN event. Yet despite these excellent observational constraints, there remains significant disagreement on the star's initial mass, with a wide range of values of $M_{\rm ZAMS}=8-20\ M_{\odot}$ (e.g., \citealt{Jencson_2023,Niu_2023,Pledger_Shara_2023,Qin_2023,Soraisam_2023,Bersten_2024,Ferrari_2024,Neustadt_2024,Singh_2024,Moriya_2024,Van_Dyk_2024,Xiang_2024}). 

Previous hydrodynamical modeling  (e.g., \citealt{Bersten_2024,Moriya_2024,Singh_2024}) pointed to low H envelope masses, and thus seemed to agree with the lower end of initial mass estimation from some progenitor studies (\citealt{Kilpatrick_2023,Neustadt_2024,Van_Dyk_2024,Xiang_2024}). In contrast, our detailed light-curve modeling indicates that SN~2023ixf does not need to be an energetic explosion of a low-mass progenitor. Instead, a more massive RSG progenitor ($M_{\rm ZAMS}\approx17-21\ M_{\odot}$) that lost more than half of its ZAMS mass can also reproduce the observed light-curve. This work is the first light-curve modeling effort that derives a progenitor mass that is consistent with the higher initial masses estimated from pre-explosion imaging (\citealt{Jencson_2023,Niu_2023, Qin_2023,Soraisam_2023}), and is further supported by the independent constraint of the progenitor star's pulsation period.

Both these options---a lower initial mass with less mass loss, as compared to a higher initial mass with more mass loss---can end up at the time of explosion with the right combination of H-envelope mass and stellar radius (see the top panel of Figure~\ref{fig:degeneracy}). This is the main reason that a range of different models can adequately approximate the light-curve shape when exploded with the appropriate explosion energy. 

It is important to recognize that these model differences are artificial, because they simply depend on the adopted mass-loss prescription, none of which resemble the much lower mass-loss rates of actual RSG winds (e.g., \citealt{Beasor_2020,Beasor_2023,Antoniadis_2024,Decin_2024}). With more realistic mass-loss rates, single RSGs at all initial masses lose very little mass via steady winds during the RSG phase, so single RSGs should retain most of their H envelopes until the time of explosion \citep{Beasor_2021}. Additionally, as noted by \citet{Beasor_2020}, the baseline ``Dutch'' mass-loss prescription overestimates normal RSG winds by a factor of $\sim$10, whereas the winds with $\eta_{\rm wind}$ values of 1.5-2.5 are even more artificially inflated. Nevertheless, the models with extremely strong winds seem to do a good job of yielding a final envelope mass that explains the light-curve shape of SN~2023ixf. 

\subsection{A Binary Companion Stripping the Envelope?}
\label{sec:binary}

In the context of RSG evolution, the most plausible way to interpret the agreement of observations with our models that have artificially inflated mass-loss rates is not that the progenitor had a bizarrely strong wind (even though this is how the models are engineered to lose their envelopes). Rather, a much more likely scenario is that the progenitor was in a binary system, and the removal of much of the H envelope was accomplished via stripping by close encounters with its companion star. If the progenitor did in fact have a higher initial mass, as our analysis of the progenitor's pulsation period seems to require, then strong binary mass stripping is essential to achieve the low envelope mass required to explain SN~2023ixf. This is consistent with the growing observational consensus that most massive stars live in binaries (e.g., \citealt{Sana_2012,de_Mink_2013,Moe_2017}) and the lack of bright precursor outbursts before SN~2023ixf (\citealt{Dong_2023,Hiramatsu_2023,Neustadt_2024,Ransome_2024}). Some studies find that binary interaction may be the formation channel for only a modest fraction of all SNe IIP (e.g., \citealt{Sravan_2019, Ercolino_2024}), which would make SN~2023ixf unusual.

Our study requires that a large mass of order 10 $M_{\odot}$ was removed from the star prior in its evolution in order to attain the appropriate properties for the observed pulsation period (see Figure~\ref{fig:periods}). For either Case A or Case B mass transfer, multiple studies predict that the donor stars in these scenarios inevitably end up as either yellow supergiants or blue supergiants, if not completely stripped to become Wolf-Rayet stars (e.g., \citealt{Gotberg_2017,Laplace_2020,Sen_2022,Marchant_2023}). This would seem at odds with the identification of SN~2023ixf's progenitor as a RSG from pre-explosion imaging (e.g., \citealt{Jencson_2023,Soraisam_2023,Van_Dyk_2024}), unless the progenitor swelled up to become red in its final years. 

Instead, a more suitable scenario for SN~2023ixf is a binary system that has undergone Case C or Case BC mass transfer, which is consistent with simulations of interacting wide massive binaries with a high initial mass ratio (\citealt{Ercolino_2024}). We caution, however, that the presence of a nearby companion close to explosion (i.e., the explosion took place during binary interaction or soon after) may modify the geometry of the progenitor, which may violate the spherical symmetry assumed when calculating the pulsation period using {\tt GYRE}. A more detailed analysis of the pulsation period for a non-spherical star may then be necessary to further constrain the progenitor mass. Nevertheless, even without the aid of pulsation period, models with high initial mass and strong envelope mass loss are still contenders for the progenitor of SN~2023ixf, as indicated by our light-curve comparisons in Section~\ref{sec:hiramatsu_models} and Section~\ref{sec:rsg_mesa_gyre}. Case BC mass transfer also paints a more self-consistent picture for the dense CSM in the vicinity of the progenitor, where most of the envelope mass is stripped during core He burning and the CSM is formed during later interactions. \citet{Matsuoka_Sawada_2024} find mass-loss rates achieved via binary interactions during the final evolutionary stages of RSG stars that are comparable to values inferred for SN~2023ixf.

Spectropolarimetric data from \cite{Vasylyev_2023}, \cite{Singh_2024}, and \cite{Shrestha_2024c} also revealed a continuum polarization level of $\approx1\%$ one day after discovery, before dropping to $\approx0.5\%$ and eventually disappearing along with narrow emission-line features over the first $\sim5-10$ days of SN~2023ixf's evolution. This indicates that either the pre-explosion mass-loss was highly asymmetric in nature (which is in agreement with the results of high-resolution spectroscopy in the first week after explosion in \citealt{Smith_2023}), the shock broke out aspherically (\citealt{Matzner_2013,Goldberg_2022b,Singh_2024}), or both. A disk-like CSM geometry is different from the spherically symmetric mass loss assumed in most models for SN~2023ixf in the literature. It does, however, seem in good agreement with the hypothesis that the progenitor was in an interacting binary system, as noted above. The motion of a nearby companion star, or perhaps a companion embedded in the common envelope of the inflated RSG progenitor, could drive mass ejection concentrated toward the equatorial plane leading up to the time of explosion (e.g., \citealt{Smith_Arnett_2014,Pejcha_2016}). In this framework, a CSM structure that consists of a dense, asymmetric disk or torus in conjunction with a low-density, extended wind (\citealt{Smith_2023,Vasylyev_2023,Singh_2024}) may be the most plausible configuration for SN~2023ixf.

\subsection{Other Possible Mechanisms for Envelope Stripping and CSM Formation}
\label{sec:other_mass_losss}

Can some other mass-loss mechanism besides binary mass stripping account for the loss of most of SN~2023ixf's H envelope? Currently, there is no observational evidence for any bright outbursts in SN~2023ixf's progenitor star for the last $\sim15$ years (\citealt{Dong_2023,Hiramatsu_2023,Neustadt_2024,Ransome_2024}). More importantly, in the case of SN~2023ixf, brief signatures of shock interaction limit the CSM to a total mass of $\lesssim0.1\ M_{\odot}$ (\citealt{Bostroem_2023b,Jacobson-Galan_2023}), lost in the few years just before core collapse, with a much weaker preceding stellar wind (\citealt{Bostroem_2024}). The dense CSM has nowhere near enough mass to account for the $\sim10M_\odot$ lost during the star's life, so the removal of the H envelope must have occurred much earlier in the progenitor's evolution. This rules out any eruptive mass-loss mechanism tied to instabilities in the final nuclear burning phases (Ne, O, and Si burning; \citealt{Arnett_Meakin_2011,Smith_Arnett_2014,Woosley_Heger_2015}) or wave driving during those same phases (\citealt{Quataert_Shiode_2012,Shiode_2013,Shiode_Quataert_2013,Fuller_2017,Wu_Fuller_2021,Wu_Fuller_2022}). 

There have been several suggestions for how to achieve strong RSG mass loss in the literature, involving local super-Eddington luminosities (\citealt{Cheng_2024}), pulsation-drive superwinds (\citealt{Heger_1997,Yoon_Cantiello_2010}), and turbulent instabilities (\citealt{Fuller_2024,Chiavassa_2011, Chiavassa_2024, Goldberg_2022a}) that may drive mass-loss rates of order $10^{-5}-10^{-4} \ M_{\odot}\ {\rm yr^{-1}}$. In order to strip such a significant portion of the progenitor's envelope, however, would require these mechanisms to have been sustained for a large fraction of the RSG phase. This is in conflict with observations of normal RSGs, which limit any enhanced mass-loss phases to about 1\% of the time during the post-main-sequence phase or less \citep{Beasor_Smith_2022}, and where mass-loss rates are observed to be lower for most of the RSG phase (\citealt{Beasor_2020,Beasor_2023,Antoniadis_2024,Decin_2024}). For the case of SN~2023ixf, these enhanced winds are also problematic because these mechanisms tend to scale with close proximity to the luminosity-to-mass ratio of the progenitor, which monotonically increases as the star evolves up the RSG branch, and as it continually loses more mass. One therefore does not expect these mechanisms to slow down or shut off with a significant H envelope remaining, as observations of SN~2023ixf require. Because of this, it is likely that the envelope mass-loss mechanism  was different from that which formed the CSM.

On the other hand, the dense CSM has more direct observational constraints, and is less in conflict with observations of normal RSGs, since it must have been formed in a very brief phase. Are there other mechanisms that could produce the dense CSM in the vicinity of SN~2023ixf's progenitor? While bright outbursts have not been detected, observations do not yet rule out the possibility that the progenitor had faint outbursts on shorter timescales compared to the cadence of these dedicated precursor searches (\citealt{Davies_2022,Dong_2023}). Such faint and brief outbursts may contribute to the CSM around the progenitor at the time of explosion. Pulsation-driven superwinds and turbulent mass ejections, while not likely to be the cause for the strong envelope stripping, may still be good candidates for forming the immediate CSM structures around SN~2023ixf's progenitor. When taken together, the combination of pulsation-driven superwinds and convectively-levitated envelope materials have also been proposed as a way to create a long-lived ``effervescent zone" around the progenitor star (\citealt{Soker_2021,Soker_2023}). A stagnant effervescent zone may, however, be in tension with the high outflow velocity in the CSM of SN~2023ixf (\citealt{Smith_2023}). 

Follow-up observations in the next decade will be crucial for determining whether any additional dense shells of CSM also exist far away from the progenitor of SN~2023ixf, as in the case of several SNe IIP that showed signatures of late-time CSM interaction years after explosion (e.g., \citealt{Maguire_2010b,Prieto_2012,Mauerhan_2017,Andrews_2018,Weil_2020}). Although difficult to prove, binary mass stripping seems to be the most straightforward and self-consistent explanation for removing most of SN~2023ixf's envelope and shaping the immediate dense CSM, and follow-up observations could similarly reveal any surviving companion. 

\section{Conclusions}
\label{sec:conclusion}

We present extensive follow-up photometric observations of SN~2023ixf in the first year of its evolution. By comparing the bolometric light curve to available grids of hydrodynamical models, we find that the plateau properties can be recovered with drastically varying physical parameters for the SN and its progenitor, owing to previously established degeneracies between explosion and progenitor properties. Motivated by this, we construct additional numerical progenitor and light-curve models. For the first time, we impose an additional constraint drawn from the observed period of the progenitor star's pre-explosion variability. Our results suggest that SN~2023ixf may have originated from the explosion of an initially massive ($M_{\rm ZAMS}\gtrsim16.5\ M_{\odot}$) RSG progenitor star, with an explosion energy of $\approx7\times10^{50}$ erg, and a $^{56}$Ni production of 0.07 $M_{\odot}$. Single-star models that agree with the observed variability of $\approx 1100$ days are characterized by a lower than usual envelope mass ($\lesssim 3\ M_{\odot}$) and a pre-explosion radius $\gtrsim 950\ R_{\odot}$. 

Currently, no isolated single-star models can explain the removal of $\sim 10\ M_{\odot}$ required to match observations of SN~2023ixf, as observed RSG winds are incapable of accomplishing such a feat. We hypothesize that the RSG progenitor likely experienced a sustained period of intense mass-loss that removed the majority of its H-rich envelope, probably due to mass stripping by a binary companion. If binary interaction was indeed the culprit that removed much of the envelope of SN~2023ixf's progenitor, it would likely be in a wide binary system with a high mass ratio in order to end up as a RSG, as theoretical models predict (\citealt{Ercolino_2024}).

Future UV observations of the SN site could help detect or constrain the presence of a surviving companion star and delineate a more complete picture of the physical origin of SN~2023ixf. With the commencement of the Vera C.~Rubin Observatory
Legacy Survey of Space and Time (LSST; \citealt{Ivezic_2019}) in 2025, high-cadence monitoring of nearby RSG populations may become increasingly common, and we will therefore have more examples of SN progenitors with known pulsation periods. The methodology presented in this work that utilizes RSG variability as an independent constraint may therefore offer an alternative way to break scaling degeneracies in future SN IIP light-curve modeling.

\acknowledgments

We thank Erez A.~Zimmerman for providing {\it Swift} streak-photometry and {\it HST} spectra that were used in earlier versions of this manuscript, Avinash Singh for supplying \ion{Fe}{2} $\lambda 5169$ line velocity measurements, Mathieu Renzo for helpful discussions and comments, and the anonymous referee for their suggestions which helped improve the quality of the manuscript. Computations in this work use High Performance Computing (HPC) resources supported by the University of Arizona TRIF, UITS, and Research, Innovation, and Impact and maintained by the UArizona Research Technologies department. We respectfully acknowledge that the University of Arizona is on the land and territories of Indigenous peoples. Today, Arizona is home to 22 federally recognized tribes, with Tucson being home to the O’odham and the Yaqui.

J.A.G~is supported by the Flatiron Research Fellowship. The Flatiron Institute is supported by the Simons Foundation. Time domain research by D.J.S.~and team is supported by NSF grants AST-1821987, 1813466, 1908972, and 2108032, and by the Heising-Simons Foundation under grant No. 20201864. K.A.B. is supported by an LSST-DA Catalyst Fellowship; this publication was thus made possible through the support of Grant 62192 from the John Templeton Foundation to LSST-DA. S.G.~is supported by an STScI Postdoctoral Fellowship. V.A.V.~acknowledges support from the NSF through grant AST-2108676. Research by Y.D.~is supported by National Science Foundation (NSF) grant AST-2008108.

L.G.~and C.P.G.~acknowledge financial support from the Spanish Ministerio de Ciencia e Innovaci\'on (MCIN), the Agencia Estatal de Investigaci\'on (AEI) 10.13039/501100011033, the European Union Next Generation EU/PRTR funds, the Horizon 2020 Research and Innovation Programme of the European Union, and by the Secretary of Universities and Research (Government of Catalonia), under the PID2023-151307NB-I00 SNNEXT project, the Marie Sk\l{}odowska-Curie and the Beatriu de Pin\'os 2021 BP 00168 programme, the 2021-SGR-01270 project, and from Centro Superior de Investigaciones Cient\'ificas (CSIC) under the PIE project 20215AT016, and the program Unidad de Excelencia Mar\'ia de Maeztu CEX2020-001058-M. The work of X.W.~is supported by the National Natural Science Foundation of China (NSFC grants 12288102 and 12033003), and the Tencent Xplorer prize.

This paper made use of data from the Las Cumbres Observatory global network of telescopes through the Supernova Key Project and Global Supernova Project. The LCO group is supported by NSF grants AST-1911151 and AST-1911225. This paper makes use of data from the AAVSO Photometric All Sky Survey, whose funding has been provided by the Robert Martin Ayers Sciences Fund and from the NSF (AST-1412587). 

Funding for SDSS-III has been provided by the Alfred P. Sloan Foundation, the Participating Institutions, the National Science Foundation, and the U.S. Department of Energy Office of Science. The SDSS-III website is \url{http://www.sdss3.org/}. SDSS-III is managed by the Astrophysical Research Consortium for the Participating Institutions of the SDSS-III Collaboration, including the University of Arizona, the Brazilian Participation Group, Brookhaven National Laboratory, Carnegie Mellon University, University of Florida, the French Participation Group, the German Participation Group, Harvard University, the Instituto de Astrofisica de Canarias, the Michigan State/Notre Dame/JINA Participation Group, Johns Hopkins University, Lawrence Berkeley National Laboratory, Max Planck Institute for Astrophysics, Max Planck Institute for Extraterrestrial Physics, New Mexico State University, New York University, Ohio State University, Pennsylvania State University, University of Portsmouth, Princeton University, the Spanish Participation Group, University of Tokyo, University of Utah, Vanderbilt University, University of Virginia, University of Washington, and Yale University.

\vspace{12pt} 

\facility{ADS, LCO (Sinistro).}

\vspace{12pt}

\textit{Software}: Astropy (\citealt{Astropy}), {\tt emcee} (\citealt{Foreman-Mackey_2013}), {\tt GYRE} (\citealt{Townsend_Teitler_2013}), {\tt lcogtsnpipe} (\citealt{Valenti_2016}), Matplotlib (\citealt{Hunter_2007}), {\tt MESA} (\citealt{Paxton_2011,Paxton_2013,Paxton_2015,Paxton_2018,Paxton_2019,Jermyn_2023}), NumPy (\citealt{Oliphant_2006}), {\tt py\_mesa\_reader} (\citealt{Wolf_2017}), SciPy (\citealt{Virtanen_2020}), {\tt STELLA} (\citealt{Blinnikov_1998,Blinnikov_2000,Blinnikov_2006,Blinnikov_2004,Baklanov_2005})

\bibliography{Reference}
\bibliographystyle{aasjournal}

\end{document}